\documentclass{svjour3}                     
\smartqed  
\usepackage[T1]{fontenc}
\usepackage[utf8]{inputenc}
\usepackage[english]{babel}

\usepackage{amssymb,amsfonts,amsmath}
\usepackage{braket}
\usepackage{graphicx}
\usepackage{subfigure}
\usepackage{hyperref}
\usepackage{enumerate}
\usepackage{multirow}
\usepackage{booktabs}
\usepackage{caption}
\usepackage{float}

\newcommand{\Tr}[0]{\operatorname{Tr}}

\newcommand{\Real}[0]{\mathbb{R}}
\newcommand{\argmin}[0]{\operatorname*{argmin}}

\begin{document}

\title{Classification Problem in a Quantum Framework
}


\author{Enrica Santucci \and Giuseppe Sergioli
}


\institute{Enrica Santucci \at
              University of Cagliari \\
             Via Is Mirrionis 1, I-09123 Cagliari, Italy\\
              \email{enrica.santucci@gmail.com}  
\and
                     Giuseppe Sergioli \at
              University of Cagliari \\
             Via Is Mirrionis 1, I-09123 Cagliari, Italy\\
              \email{giuseppe.sergioli@gmail.com}
}

\date{Received: date / Accepted: date}

\maketitle

\begin{abstract}
The aim of this paper is to provide a quantum counterpart of the well known minimum-distance classifier named {\em Nearest Mean Classifier} (NMC). In particular, we refer to the following previous works: \emph{i)} in \cite{sergioli2016} we have introduced a detailed quantum version of the NMC, named \emph{Quantum Nearest Mean Classifier} (QNMC), for two-dimensional problems and we have proposed a generalization to abitrary dimensions; \emph{ii)} in \cite{sergioli2017} the $n$-dimensional problem was analyzed in detail and a particular encoding for arbitrary $n$-feature vectors into density operators has been presented. In this paper, we introduce a new promizing encoding of arbitrary $n$-dimensional patterns into density operators, starting from the two-feature encoding provided in \cite{sergioli2016}. Further, unlike the NMC, the QNMC shows to be not invariant by rescaling the features of each pattern. This property allows us to introduce a free parameter whose variation provides, in some case, an improvement of the QNMC performance. We show experimental results where: \emph{i)} the NMC and QNMC performances are compared on different datasets; \emph{ii)} the effects of the non-invariance under uniform rescaling for the QNMC are investigated.
\keywords{Minimum Distance Classifier \and Density Operators \and Rescaling Invariance}
\end{abstract}

\section{Introduction}
\label{sec:1}

The aim of this paper is to propose a classification algorithm inspired by quantum theory. In particular, the classification process we propose consists in the following steps: $i)$ we encode a dataset of reals objects (represented by real vectors) into quantum objects (represented by density operators); $ii)$ we perform a particular classification process of quantum objects by involving a standard notion of distance between quantum states (the \emph{trace distance}); $iii)$ we naturally decode the results of the classification of quantum objects for the initial real dataset.

The result is that we can perform (by using a classical computer) a classification of real objects by involving the formalism of quantum mechanics. Interstingly enough, the error of this quantum-inspired encoding turns out to be smaller with respect to the well known nearest mean classifier for many different datasets.  

However, some attempts to represent a classification process by appealing to the quantum theory was already realized \cite{aimeur2006,12,13,Rebentrost,sergioli2017,sergioli2016}. In particular, the problem to find a \emph{more convenient} encoding from classical to quantum object is nowaday an open and interesting problem \cite{10,8}. 

In this work we propose a new encoding that  leads to three relevant advantages: \emph{i)} it allows to simply encode an arbitrary $n$-feature pattern into a quantum state; \emph{ii)} the classification process performance turns out to be better than the standard NMC for several and different kinds of datasets;  \emph{ii)} for some dataset, this new encoding exhibits a further advantage that can be gained by exploiting the non-invariance under rescaling of the quantum-inspired classifier.

The paper is organized as follows: in Section \ref{sec:NMC} we briefly describe the classification process and, in particular, the formal structure of the NMC. Section \ref{sec:encoding} is devoted to the definition of a new encoding of real patterns into quantum states. In Section \ref{sec:QNMC} we introduce the quantum version of the NMC based on the new encoding previously described. In Section \ref{sec:exp} we compare the NMC and the QNMC on different datasets showing that, in general, the QNMC exhibits better performances (in terms of accuracy and other significant statistical quantities) with respect to the NMC. Further, starting from the fact that, differently from the NMC, the QNMC is not invariant under rescaling, we also show that, for some dataset, it is possible to provide a benefit from this non-invariance property. Some conclusions and possible further developments are proposed at the end of the paper.

\section{On the classification process}
\label{sec:NMC}

Here, we address the classification problem, which is an instance of supervised learning, \emph{i.e.} learning from a training set of correctly labeled objects. More precisely, each object can be characterized by its features; hence, a $d$-feature object can be naturally represented by a $d$-dimensional real vector, i.e. $\vec x = [x^{(1)}, \ldots, x^{(d)}]\in \mathcal{X}$, where $\mathcal{X}\subseteq\Real^d$ is generally a subset of the $d$-dimensional real space and represents the \emph{feature space}. Hence, any arbitrary object is represented by a vector $\vec x$ associated to a given class of objects (but, in principle, we do not know which one). Let $\mathcal Y=\{1, \ldots, L\}$ be the class label set. A \emph{pattern} is represented by a pair $(\vec x,y)$, where  $\vec x$ is the \emph{feature vector} representing an object and $y \in\mathcal Y$ is the \emph{label} of the class which $\vec x$ is associated to. The aim of the classification process  is to identify which class $\vec x$ belongs to, by learning about the set of objects whose class is known. Then, the so called \emph{training set} is given by $\mathcal S_{\text{tr}} = \{({\bf x}_n, y_n)\}_{n=1}^N$, where ${\bf x}_n \in \mathcal X$, $y_n \in \mathcal Y$ ($\forall n=1, \ldots, N$) and $N$ is the number of patterns belonging to $\mathcal S_{\text{tr}}$. Finally, let $N_l$ be the cardinality of the training set associated to the $l$-th class (for $l=1, 2, \ldots, L$) such that $\sum_{l=1}^L N_l = N$.

We now introduce the well known \emph{Nearest Mean Classifier} (NMC) \cite{DuHa}, which is a particular kind of minimum-distance classifier widely used in pattern recognition. The strategy consists in computing the distances between a pattern $\bf x$ (to classify) and other patterns chosen as prototypes of each class (called \emph{centroids}). Finally, $\bf x$ is labeled as belonging to the  class whose distance is minimum. So, we can resume the NMC algorithm as follows:
\begin{enumerate}
  \item The computation of the \textit{centroid} (\emph{i.e.} the sample mean \cite{Johnson}) associated to each class is given by:
  \begin{equation}\label{eq:ccentroid}
  \vec \mu_l = \frac{1}{N_l} \sum_{n=1}^{N_l} \vec x_n,\quad l=1, 2, \ldots, L;
  \end{equation}
  \item The classification of the  pattern $\vec x$ is provided by:
  \begin{equation}\label{eq:cclassify}
  \argmin_{l= 1, \ldots L} d_E(\vec x, \vec \mu_l),
  \end{equation}
where $d_E(\vec x, \vec \mu_l) = \| \vec x - \vec \mu_l\|$ is the Euclidean distance between the pattern $\vec x$ and the centroid $\vec \mu_l$.
\end{enumerate}
Briefly speaking, if the pattern $\vec x$ is \emph{closest} to the centroid $\vec \mu_l$, then $\vec x$ is labeled by $l$, \emph{i.e.} $y = l$.

As a remark, it is worth noting that, depending on the particular distribution of the patterns of the dataset, it is possible that a pattern belonging to a given class is closest to the centroid of another class. In this case, if the algorithm would be applied to this pattern, it would fail.
Hence, for an arbitrary pattern $\vec x$ whose class is \textit{a priori} unknown, the output of above classification process has the following four possibilities \cite{F}:  \emph{i)\ True Positive} (TP): pattern belonging to the $l$-th class and correctly classified as $l$; \emph{ii)\ True Negative} (TN): pattern belonging to a class different than $l$, and correctly classified as not $l$; \emph{iii)\ False Positive} (FP): pattern belonging to a class different than $l$, and uncorrectly classified as $l$; \emph{iv)\ False Negative} (FN): pattern belonging to the $l$-th class, and uncorrectly classified as not $l$.

In order to evaluate the performance of a certain classification algorithm, the standard procedure consists in dividing the original labeled dataset $\mathcal S$ of $N'$ patterns, into a training set $\mathcal{S}_{\text{tr}}$ of $N$ patterns and a set $\mathcal{S}_{\text{ts}}$ of $(N' - N)$ patterns (\emph{i.e.} $\mathcal{S} = \mathcal{S}_{\text{tr}} \cup \mathcal{S}_{\text{ts}}$). This set $\mathcal{S}_{\text{ts}}$ of patterns is called \emph{test set} \cite{DuHa} and it is defined as $\mathcal{S}_{\text{ts}} = \{({\bf x}_n, y_n)\}_{n=N+1}^{N'}$, where $N'_l$ (such that $\sum_{l=1}^L N'_l = N' - N$) is the cardinality of the test set associated to the $l$-th class. 

Then, by applying the NMC to the test set, it is possible to evaluate the classification algorithm performance by considering the following statistical measures associated to each class $l$ depending on the quantities listed above:
\begin{itemize}
\item \emph{True Positive Rate} (TPR): $\text{TPR}= \frac{\text{TP}}{\text{TP} + \text{FN}}$;
\item \emph{True Negative Rate} (TNR): $\text{TNR}= \frac{\text{TN}}{\text{TN} + \text{FP}}$;
\item \emph{False Positive Rate} (FPR): $\text{FPR}= \frac{\text{FP}}{\text{FP} + \text{TN}}= 1- \text{TPN}$;
\item \emph{False Negative Rate} (FNR): $\text{FNR}= \frac{\text{FN}}{\text{FN} + \text{TP}}=1 -\text{TPR}$.
\end{itemize}
Further, other standard statistical indeces \cite{F} used to establish the reliability of a classification algorithm are:
\begin{itemize}
\item \emph{Classification error} (E): $\text{E} = 1 - \frac{\text{TP}}{\text{N}' - \text{N}}$;
 \item \emph{Precision} (P): $\text{P} = \frac{\text{TP}}{\text{TP} + \text{FP}}$;
\item \emph{k's Cohen} (K): $\text{K} = \frac{\text{Pr(a)} - \text{Pr(e)}}{1 - \text{Pr(e)}}$, where \\$\text{Pr(a)} = \frac{\text{TP} + \text{TN}}{\text{N}' - \text{N}}$,\ $\text{Pr(e)} = \frac{(\text{TP} + \text{FP})(\text{TP} + \text{FN}) + (\text{FP} + \text{TN})(\text{TN} + \text{FN})}{(\text{N}' - \text{N})^2}$.
\end{itemize}

In particular, the classification error represents the percentage of misclassified patterns, the precision is a measure of the statistical variability of the considered model and the k's Cohen represents the degree of agreement among items that can assume values ranging from $-1$ to $+1$ (K $=+1$ corresponds to a perfect classification procedure while K $=-1$ corresponds to a  completely wrong classification).
Let us note that these statistical parameters have to be a part considered for each class. Then, the final value of each statistical parameter related to the classification algorithm is the weighted sum of the statistical parameters of each class.

\section{Correspondence between pattern and density operator}
\label{sec:encoding}
In order to introduce a quantum version of the NMC, the first step is to find an appropriate quantum encoding for a real pattern.

Generally, given a $d$-dimensional feature vector, there exist different ways to encode it into a density operator \cite{8}. In \cite{sergioli2016} the following encoding was introduced. Let us consider the inverse of the steregraphic projection \cite{Cox} given by:
\begin{equation}\label{sp1}
SP^{-1}:[x^{(1)},x^{(2)}] \mapsto \Big [\frac{2x^{(1)}}{||x||^2+1},\frac{2x^{(2)}}{||x||^2+1},\frac{||x||^2-1}{||x||^2+1}\Big ],
\end{equation}
where $||x||^2 = [x^{(1)}]^2 + [x^{(2)}]^2$.
Then, by imposing that $r_1 = \frac{2x^{(1)}}{||x||^2 + 1},\ r_2 = \frac{2x^{(2)}}{||x||^2 + 1},\ r_3 = \frac{||x||^2 - 1}{||x||^2 + 1}$, if we consider $r_1, r_2, r_3$ as Pauli components\footnote{We consider the representation of an arbitrary density operator as linear combination of Pauli matrices.} of a density operator $\rho_{\bf x}\in\mathbb C^2$, then the density operator associated to the pattern $\vec x = [x^{(1)},x^{(2)}]$ can be written as:
\begin{equation}\label{eqn:desnity-pattern}
\frac{1}{2}\begin{pmatrix}1+r_3 & r_1-ir_2 \\
          r_1+ir_2 & 1-r_3  
\end{pmatrix}=\frac{1}{||x||^2+1}
\begin{pmatrix}
	||x||^2 & x^{(1)}-ix^{(2)} \\
	x^{(1)}+ix^{(2)} & 1  
\end{pmatrix}.
\end{equation}
The advantage in using this encoding consists in the fact that it provides an easy visualization of an arbitrary two-feature vector on the Bloch sphere \cite{sergioli2016}. However, the main problem regards the generalization for $d$-feature vectors. Although in \cite{sergioli2016} a generalization to the $d$-feature case was introduced, it exhibits some difficulties to be implemented for general cases.

An alternative  encoding of a $d$-feature pattern $\vec x = [x^{(1)}, \ldots, x^{(d)}]$ into a density operator was proposed in \cite{sergioli2017}.
It is obtained {\emph{i)}} by mapping $\vec x\in\mathbb{R}^d$ into a ($d+1$)-dimensional vector $\vec x'\in\mathbb{R}^{d+1}$ achieved by the generalized Eq. (\ref{sp1}), \emph{i.e.}
\begin{equation}\label{eq:sp2}
  \vec x' = SP^{-1}(\vec x) = \frac{1}{||\vec x||^2+1} \left[2x^{(1)}, \ldots, 2x^{(d)}, ||\vec x||^2 - 1 \right]
\end{equation}
where $||\vec x||^2 = \sum_{i=1}^d [x^{(i)}]^2$, and then {\emph{ii)}} by considering the projector $\rho_{\vec x} = \vec x'\cdot (\vec x')^T$.

In this work we propose a different version of the QNMC based on a new  encoding again and we show that this exhibits interesting improvements mostly by exploiting the non invariance under rescaling of the features.

Accordingly with \cite{12,Rebentrost,8}, when a real vector is encoded into a quantum state, in order to avoid a loss of information it is important that the quantum state keeps some information about the norm of the original real vector. In light of this fact, we introduce the following alternative encoding.

Let $\vec x = [x^{(1)}, \ldots, x^{(d)}]\in \mathbb{R}^d$ be an arbitrary $d$-feature vector.
\begin{enumerate}
\item We maps the vector $\vec x \in \mathbb{R}^d$ into a vector $\vec x' \in \mathbb{R}^{d+1}$, whose first $d$ features are the components of the vector $\vec x$ and the $(d+1)$-th feature is the norm of $\vec x$. Formally:
\begin{equation}
\vec x = [x^{(1)}, \ldots, x^{(d)}]\ \mapsto\ \vec x' = [x^{(1)}, \ldots, x^{(d)}, ||\vec x||].
\end{equation}
\item Finally, we obtain the vector $\vec x'$ by dividing the first $d$ components of the vector $\vec x'$ for $||\vec x||$: 
\begin{equation}
\vec x' \ \mapsto\ \vec x'' = \Big [\frac{x^{(1)}}{||\vec x||}, \ldots, \frac{x^{(d)}}{||\vec x||}, ||\vec x||\Big ].
\end{equation}
\item We consider the norm of the vector $\vec x''$, \emph{i.e.} $||\vec x''|| = \sqrt{||\vec x||^2 + 1}$ and we map the vector $\vec x''$ into the normalized vector $\vec x'''$ as follows: 
\begin{equation}\label{x'''}
\vec x'' \ \mapsto\ \vec x''' = \frac{\vec x''}{||\vec x''||}= \Big [\frac{x^{(1)}}{||\vec x||\sqrt{||\vec x||^2 + 1}}, \ldots, \frac{x^{(d)}}{||\vec x||\sqrt{||\vec x||^2 + 1}}, \frac{||\vec x||}{\sqrt{||\vec x||^2 + 1}}\Big ].
\end{equation}
\end{enumerate}
Now, we provide the following definition.
\begin{definition}[Density pattern]
\label{def:dp}

Let $\vec x = [x^{(1)}, \ldots, x^{(d)}]$ be an arbitrary $d$-features pattern. Then, the matrix representation of the \emph{density pattern} $\rho_{\vec x}$ corresponding to the pattern $\vec x$ is defined as:
\begin{equation}
 \label{eq:dp}
\rho_{\vec x} \doteq \vec x'''\cdot(\vec x''')^\dagger
\end{equation}
where $\vec x'''$ is given by Eq (\ref{x'''}). 
\end{definition}
Hence, this encoding maps real $d$-dimensional vectors ${\vec x}$ into $(d+1)$-dimensional pure states $\rho_{\vec x}$. In this way, we obtain an encoding that takes into account the information about the initial real vector norm and, in the meantime, allows to easily encode also arbitrary $d$-dimensional real patterns.

\section{Quantum classification}
\label{sec:QNMC}

In this section we introduce a quantum-inspired version of the NMC, named \textit{Quantum Nearest Mean Classifier} (QNMC). It can be seen as a particular kind of minimum-distance classifier between quantum objects (\emph{i.e.} density patterns). The use of this new formalism could lead not only to achieve the well known advantages related to the quantum computation with respect to the classical one (mostly related to the speed up of the computation process), but also to make a full comparison between NMC and QNMC performance by using a classical computer only.

In order to provide a quantum counterpart of the NMC, we need: \emph{i)} an encoding from real patterns to quantum objects (already defined in the previous section); \emph{ii)} a quantum counterpart of the classical centroid (\emph{i.e.} a sort of class prototype), that will be named \textit{quantum centroid}; \emph{iii)} a suitable definition of \textit{quantum distance} between density patterns, that plays the same role as the Euclidean distance for the NMC.
In this quantum framework, the quantum version $\mathcal S^q$ of the dataset $\mathcal S$ is given by:
$$ \mathcal S^q = \mathcal S^q_{\text{tr}} \cup \mathcal S^q_{\text{ts}}, \quad
 \mathcal S^q_{\text{tr}} = \{(\rho_{\vec x_n}, y_n)\}_{n=1}^N, \quad \mathcal S^q_{\text{ts}} = \{(\rho_{\vec x_n}, y_n)\}_{n=N+1}^{N'},$$
where $\rho_{\vec x_n}$ is the density pattern associated to the pattern $\vec x_n$ and $y_n$ is its original label.
Then, $\mathcal S^q_{\text{tr}}$ and $\mathcal S^q_{\text{ts}}$ represent the quantum versions of training and test set respectively, \emph{i.e.} the sets of all the density patterns obtained by encoding all the elements of $\mathcal S_{\text{tr}}$ and $\mathcal S_{\text{ts}}$. Now, we naturally introduce the quantum version of the classical centroid $\vec \mu_l$, given in Eq.~\eqref{eq:ccentroid}, as follows.
\begin{definition}[Quantum centroid]
\label{def:qcentroid}
Let $\mathcal S^q$ be a labeled dataset of $N'$ density patterns such that $\mathcal S^q_{\text{tr}} \subseteq \mathcal S^q$ is a training set composed of $N$ density patterns. Further, let $\mathcal Y = \{1,2, \ldots, L\}$ be the set of class labels.
The \emph{quantum centroid} of the $l$-th class is given by:
\begin{equation}
\label{eq:qcentroid}
\rho_{l} = \frac{1}{N_l} \sum_{n=1}^{N_l}\rho_{\vec x_n},\quad l=1, \ldots, L
\end{equation}
where $N_l$ is the number of density patterns of the $l$-th class belonging to $\mathcal S^q_{\text{tr}}$, such that $\sum_{l=1}^L N_l = N$.
\end{definition}
Notice that the quantum centroids are generally mixed states and they are not obtained by encoding the classical centroids $\vec \mu_l$, \emph{i.e.} 
\begin{equation}
\rho_{l} \neq \rho_{\vec \mu_l},\ \forall l=1, \ldots,L.
\end{equation}
Accordingly, the definition of the quantum centroid leads to a new object that is no longer a pure state and has not any classical counterpart. This is the main reason that establishes, even in a foundamental level, the difference between NMC and QNMC. In particular, it is easy to verify \cite{sergioli2016} that, unlike the classical case, the expression of the quantum centroid is sensitive to the dataset dispersion.

In order to consider a suitable definition of distance between density patterns, we recall the well known definition of trace distance between quantum states (see, \emph{e.g.}~\cite{nielsenbook}).
\begin{definition}[Trace distance]
\label{def:trdist}
Let $\rho$ and $\rho'$ be two quantum density operators belonging to the same dimensional Hilbert space. The \emph{trace distance} between them is given by:
\begin{equation}\label{eq:trdist}
  d_T(\rho,\rho') = \frac{1}{2} \Tr|\rho - \rho'|,
\end{equation}
where $|A| = \sqrt{A^\dag A}$.
\end{definition}
Notice that the trace distance is a true metric for the density operators, that is, it satisfies: \emph{i)} $d_T(\rho,\rho') \geq 0$ with equality iff $\rho=\rho'$ (\emph{positivity}), \emph{ii)}   $d_T(\rho,\rho') = d_T(\rho',\rho)$ (\emph{symmetry}) and \emph{iii)} $d_T(\rho,\rho')+d_T(\rho',\rho'') \geq d_T(\rho,\rho'')$ (\emph{triangle inequality}).

We have introduced all the ingredients we need to describe the QNMC process, that, similarly to the classical case, consists in the following steps:
\begin{itemize}
  \item to construct quantum training and test sets $\mathcal S^q_{\text{tr}}$, $\mathcal S^q_{\text{ts}}$ by applying the encoding introduced in Definition~\ref{def:dp} to each pattern of the classical training and test sets $\mathcal S_{\text{tr}}$, $\mathcal S_{\text{ts}}$;
  \item to calculate the quantum centroids $\rho_l$ ($\forall l=1, \ldots L$), by using the quantum training set $\mathcal S^q_{\text{tr}}$, according to Definition~\ref{def:qcentroid};
  \item to classify an arbitrary density pattern $\rho_{\vec x}\in S^q_{\text{ts}}$ accordingly with the following minimization problem:
 \begin{equation}\label{eq:cclassify}
 \argmin_{l= 1,\ldots,L} d_T(\rho_{\vec x}, \rho_l),
 \end{equation}
where $d_T$ is the Trace distance introduced in Definition~\ref{def:trdist}.
\end{itemize}

\section{Experimental results}
\label{sec:exp}

This section is devoted to show a comparison between the NMC and the QNMC performances in terms of the statistical parameters introduced in Section \ref{sec:NMC}. We use both classifiers to analyze fourteen datasets. In particular, two different kinds of datasets have been studied: five of them (\emph{Gaussian (I), Gaussian (II), Gaussian (III), Moon, Banana}) are artificial datasets (in particular, the first three datasets follow Gaussian distributions), while the others (\emph{Balance, Bands, Breast Cancer (I), Breast Cancer (II), Ilpd, Ionosphere, Liver, Pima, Tic Tac}) are real-world datasets, extracted from the UCI repository\footnote{\url{http://archive.ics.uci.edu/ml}} and following unknown distributions. We stress that, in real situations, we usually deal with datasets following unknown distributions, then the most interesting case is the second one. However, the use of artificial datasets following known distribution, and in particular Gaussian distributions with specific parameters, can help to catch precious information that will be discussed in the next section. 

\subsection{Comparison between QNMC and NMC}
\label{sec:comparison}
In Table \ref{tab:1} we summarize the characteristics of the datasets involved in our experiments. In particular, for each dataset we list the total number of patterns, the number of patterns belonging to each class and the number of features. Let us note that, although we mostly confine our investigation to two-class datasets, our model can be easily extended -without any loss of generality- to multiclass problems (as we show for the three-class datasets \emph{Balance} and \emph{Gaussian (III)}).

In order to make our results statistically significant, we apply the standard procedure of splitting each dataset in training and test sets composed of the $\%80$ and $\%20$ of the total patterns respectively, and we carry out ten experiments for each dataset, where the splitting is every time randomly taken.

In Table \ref{tab:2}, we report QNMC and NMC performance for each dataset, evaluated in terms of mean value and standard deviation (computed on ten runs) of the statistical indexes, discussed in the previous section. For the sake of semplicity, we omit the values of FPR and FNR because they can be easily obtained by TPR and TNR values (\emph{i.e.} FPR = 1 - TNR, FNR = 1 - TPR).

\begin{table}[tb]
\caption{Characteristics of the datasets used in our experiments. The number of instances in each class is shown between brackets.}
\label{tab:1}
\centering
\begin{tabular}{llr}
\hline\noalign{\smallskip}
\textbf{Data set}  & {\bf Instances} & {\bf Features} ($d$) \\
\hline\noalign{\smallskip}
Balance & 625 (49+288+288) & 4 \\ 
Banana & 5300 (2376+2924) & 2 \\ 
Bands & 365 (135+230) & 19 \\ 
Breast Cancer (I) & 683 (444+239) & 10 \\ 
Breast Cancer (II) & 699 (458+241) & 9 \\ 
Ilpd & 583 (416+167) & 9 \\ 
Ionosphere & 351 (225+126) & 34 \\ 
Liver & 578 (413+165) & 10 \\ 
Moon & 200 (100+100) & 2 \\ 
Pima  & 768 (500+268) & 8 \\ 
TicTac  & 958 (626+332) & 9 \\ 
Gaussian (I)  & 400 (200+200) & 30 \\ 
Gaussian (II)  & 1000 (100+900) & 8 \\ 
Gaussian (III)  & 2050 (50+500+1500) & 8 \\
\noalign{\smallskip}\hline\noalign{\smallskip}
\end{tabular}
\end{table}

\begin{center}
\begin{table}[tb]
\resizebox{0.9\textwidth}{!}{\begin{minipage}{\textwidth}
\caption{Comparison between QNMC and NMC performances.}
\label{Nmc}
\centering
\begin{tabular}{llllll r|r|r|r|}
\hline\noalign{\smallskip}
\multicolumn{1}{c}{ }  &  \multicolumn{ 5}{c}{ {\bf QNMC}} \\ 
\hline\noalign{\smallskip}
\multicolumn{1}{c}{ {\bf Dataset}} &  \multicolumn{1}{c}{E} & \multicolumn{1}{c}{TPR} & \multicolumn{1}{c}{TNR} & \multicolumn{1}{c}{P} & \multicolumn{1}{c}{K}    \\ 
\hline\noalign{\smallskip}
	Balance   & 0.148 $\pm$ 0.018 & 0.852 $\pm$ 0.018 & 0.915 $\pm$ 0.014 & 0.862 $\pm$ 0.022 & 0.767 $\pm$ 0.029 \\ 
	Banana   &  0.316 $\pm$ 0.017 & 0.684 $\pm$ 0.017 & 0.660 $\pm$ 0.017 & 0.684 $\pm$ 0.018 & 0.350 $\pm$ 0.034 \\ 
	Bands   & 0.394 $\pm$ 0.053 & 0.606 $\pm$ 0.053 & 0.528 $\pm$ 0.071 & 0.606 $\pm$ 0.058 & 0.133 $\pm$ 0.112 \\ 
	Breast Cancer (I) & 0.386 $\pm$ 0.038 & 0.614 $\pm$ 0.038 & 0.444 $\pm$ 0.045 & 0.583 $\pm$ 0.044 & 0.062 $\pm$ 0.069 \\ 
	Breast Cancer (II) & 0.040 $\pm$ 0.015 & 0.946 $\pm$ 0.023 & 0.986 $\pm$ 0.016 & 0.993 $\pm$ 0.009 & 0.912 $\pm$ 0.033\\ 
	Ilpd   & 0.351 $\pm$ 0.037 & 0.649 $\pm$ 0.037 & 0.705 $\pm$ 0.056 & 0.734 $\pm$ 0.041 & 0.292 $\pm$ 0.073 \\ 
	Ionosphere  & 0.165 $\pm$ 0.049 & 0.835 $\pm$ 0.049 & 0.764 $\pm$ 0.059 & 0.842 $\pm$ 0.051 & 0.624 $\pm$ 0.105 \\ 
	Liver   & 0.342 $\pm$ 0.037 & 0.607 $\pm$ 0.057 & 0.783 $\pm$ 0.059 & 0.870 $\pm$ 0.039 & 0.318 $\pm$ 0.061\\ 
	Moon   & 0.156 $\pm$ 0.042 & 0.857 $\pm$ 0.063 & 0.831 $\pm$ 0.066 & 0.841 $\pm$ 0.066 & 0.683 $\pm$ 0.085 \\ 
	Pima   & 0.304 $\pm$ 0.030 & 0.696 $\pm$ 0.030 & 0.690 $\pm$ 0.044 & 0.720 $\pm$ 0.030 & 0.365 $\pm$ 0.066 \\ 
	Tic Tac  & 0.410 $\pm$ 0.032 & 0.590 $\pm$ 0.032 & 0.597 $\pm$ 0.039 & 0.629 $\pm$ 0.036 & 0.172 $\pm$ 0.061 \\
	Gaussian (I)  & 0.274 $\pm$ 0.051 & 0.726 $\pm$ 0.051 & 0.728 $\pm$ 0.049 & 0.745 $\pm$ 0.048 & 0.452 $\pm$ 0.099  \\ 
	Gaussian (II)  & 0.210 $\pm$ 0.025 & 0.790 $\pm$ 0.025 & 0.744 $\pm$ 0.061 & 0.900 $\pm$ 0.019 & 0.308 $\pm$ 0.058 \\ 
        Gaussian (III)  & 0.401 $\pm$ 0.036 & 0.599 $\pm$ 0.036 & 0.558 $\pm$ 0.026 & 0.654 $\pm$ 0.041 & 0152 $\pm$ 0.043 \\
\noalign{\smallskip}\hline\noalign{\smallskip}
\multicolumn{1}{c}{ }  &  \multicolumn{ 5}{c}{ {\bf NMC}} \\ 
\hline\noalign{\smallskip}
\multicolumn{1}{c}{ {\bf Dataset}} &  \multicolumn{1}{c}{E} & \multicolumn{1}{c}{TPR} & \multicolumn{1}{c}{TNR} & \multicolumn{1}{c}{P} & \multicolumn{1}{c}{K}    \\ 
\hline\noalign{\smallskip}
	Balance   & 0.267 $\pm$ 0.038 & 0.733 $\pm$ 0.038 & 0.969 $\pm$ 0.014 & 0.925 $\pm$ 0.025 & 0.686 $\pm$ 0.034 \\ 
	Banana   & 0.453 $\pm$ 0.019 & 0.548 $\pm$ 0.019 & 0.552 $\pm$ 0.020 & 0.556 $\pm$ 0.020 & 0.098 $\pm$ 0.038 \\ 
	Bands   & 0.435 $\pm$ 0.048 & 0.565 $\pm$ 0.048 & 0.582 $\pm$ 0.055 & 0.605 $\pm$ 0.054 & 0.135 $\pm$ 0.092 \\ 
	Breast Cancer (I) & 0.442 $\pm$ 0.037 & 0.558 $\pm$ 0.037 & 0.464 $\pm$ 0.046 & 0.551 $\pm$ 0.039 & 0.022 $\pm$ 0.076 \\ 
	Breast Cancer (II) & 0.042 $\pm$ 0.015 & 0.973 $\pm$ 0.015 & 0.931 $\pm$ 0.032 & 0.963 $\pm$ 0.017 & 0.908 $\pm$ 0.033 \\ 
	Ilpd   & 0.470 $\pm$ 0.037 & 0.530 $\pm$ 0.037 & 0.757 $\pm$ 0.041 & 0.761 $\pm$ 0.037 & 0.193 $\pm$ 0.051 \\ 
	Ionosphere  & 0.323 $\pm$ 0.051 & 0.677 $\pm$ 0.051 & 0.676 $\pm$ 0.051 & 0.680 $\pm$ 0.051 & 0.351 $\pm$ 0.102 \\ 
	Liver   & 0.472 $\pm$ 0.048 & 0.388 $\pm$ 0.057 & 0.891 $\pm$ 0.055 & 0.905 $\pm$ 0.045 & 0.193 $\pm$ 0.060\\ 
	Moon   & 0.234 $\pm$ 0.065 & 0.772 $\pm$ 0.089 & 0.762 $\pm$ 0.085 & 0.771 $\pm$ 0.091 & 0.528 $\pm$ 0.130\\ 
	Pima   & 0.375 $\pm$ 0.033 & 0.625 $\pm$ 0.033 & 0.546 $\pm$ 0.045 & 0.622 $\pm$ 0.037 & 0.173 $\pm$ 0.075 \\ 
	Tic Tac  & 0.439 $\pm$ 0.031 & 0.561 $\pm$ 0.031 & 0.571 $\pm$ 0.042 & 0.606 $\pm$ 0.036 & 0.119 $\pm$ 0.063 \\ 
	Gaussian (I)  & 0.322 $\pm$ 0.042 & 0.679 $\pm$ 0.042 & 0.680 $\pm$ 0.043 & 0.685 $\pm$ 0.042 & 0.355 $\pm$ 0.085  \\ 
	Gaussian (II) & 0.320 $\pm$ 0.032 & 0.680 $\pm$ 0.032 & 0.588 $\pm$ 0.102 & 0.860 $\pm$ 0.032 & 0.129 $\pm$ 0.055 \\ 
        Gaussian (III)  & 0.530 $\pm$ 0.029 & 0.470 $\pm$ 0.029 & 0.625 $\pm$ 0.030 & 0.620 $\pm$ 0.036 & 0.066 $\pm$ 0.044 \\ 
\noalign{\smallskip}\hline\noalign{\smallskip}
\end{tabular}
\label{tab:2}
\end{minipage} }
\end{table}
\end{center}
We observe, by comparing QNMC and NMC performances (see Table \ref{tab:2}), that the first provides a significant improvement with respect to the standard NMC in terms of all the statistical parameters we have considered. Further, the new encoding, for two-feature datasets, provides better performance than the one considered in \cite{sergioli2016} (where the QNMC error with related standard deviation was $0.174 \pm 0.047$ for \emph{Moon} and $0.419 \pm 0.015$ for \emph{Banana}) and it generally exhibits quite similar performance with respect to the one in \cite{sergioli2017} for multi-dimension datasets, except in the case of \emph{Breast Cancer (II)} and \emph{Gaussian (I)} datasets, for which the new encoding provides a classification improvement of about $3\%$ and $5\%$, respectively.

The artificial Gaussian datasets may deserve a brief comment. Let us discuss the way in which the three Gaussian datasets have been created. The first one, called \emph{Gaussian (I)} \cite{Skurichina02} is a perfectly balanced dataset (\emph{i.e.} both classes have the same number of patterns), patterns have the same dispersion in both classes, and only some features are correlated \cite{Wass}. The second one, called \emph{Gaussian (II)}, is an unbalanced dataset (\emph{i.e.} classes have a very different number of patterns), patterns have not the same dispersion in both classes and features are not correlated. Finally, the third one, called \emph{Gaussian (III)}, is composed of three classes and it is an unbalanced dataset with different pattern dispersion in all the classes, where all the features are correlated. 

For these Gaussian datasets, the NMC is not the best classifier \cite{DuHa} because of the particular characteristics of the class dispersion. Indeed, the NMC it does not take into account of the data dispersion. Conversely, by looking at Table \ref{tab:2}, the improvements of the QNMC seems to exhibit some kind of sensitivity of the classifier with respect to the data dispersion. A detailed description of this problem will be addressed in a future work.

As a remark, it is important to remind that, even if it is possible to establish which is a good or bad classifier for a given dataset by the evaluation of some a priori data characteristics, generally it is no possible to establish an absolute superiority of a given classifier for any dataset, according to the well known \emph{No Free Lunch Theorem} \cite{DuHa}. Anyway, the QNMC seems to be particularly convenient when the data distribution is difficult to treat with the standard NMC.
\subsection{Non-invariance under rescaling}
\label{sec:noninvariance}

The final experimental results that we present in this paper regard a significant difference between NMC and QNMC. Let us suppose that all the features of the patterns $\vec x_n$ ($\forall n=1, \ldots, N'$) belonging to the original dataset $\mathcal S$ are multiplied by the same parameter $t\in \mathbb{R}$, \emph{i.e.} $\vec x_n \mapsto t\vec x_n$. Then, the whole dataset is subjected to an increasing dispersion (for $|t| > 1$) or a decreasing dispersion (for $|t| < 1$) and the classical centroids change according to $\vec \mu_l \mapsto t\vec \mu_l$ ($\forall l=1, \ldots, L$). Consequently, the classification problem for each pattern of the rescaled test set can be written as $$\argmin_{l=1,\ldots,L} d_E(t \vec x_{n}, t \vec \mu_l) = t  \argmin_{l=1,\ldots,L} d_E(\vec x_{n},\vec \mu_l),\quad \forall n=N+1, \ldots,N'.$$

\begin{figure}[ht!]
\centering
\subfigure[]{\label{fig:Ionosphere}
\includegraphics[width=0.484\columnwidth]{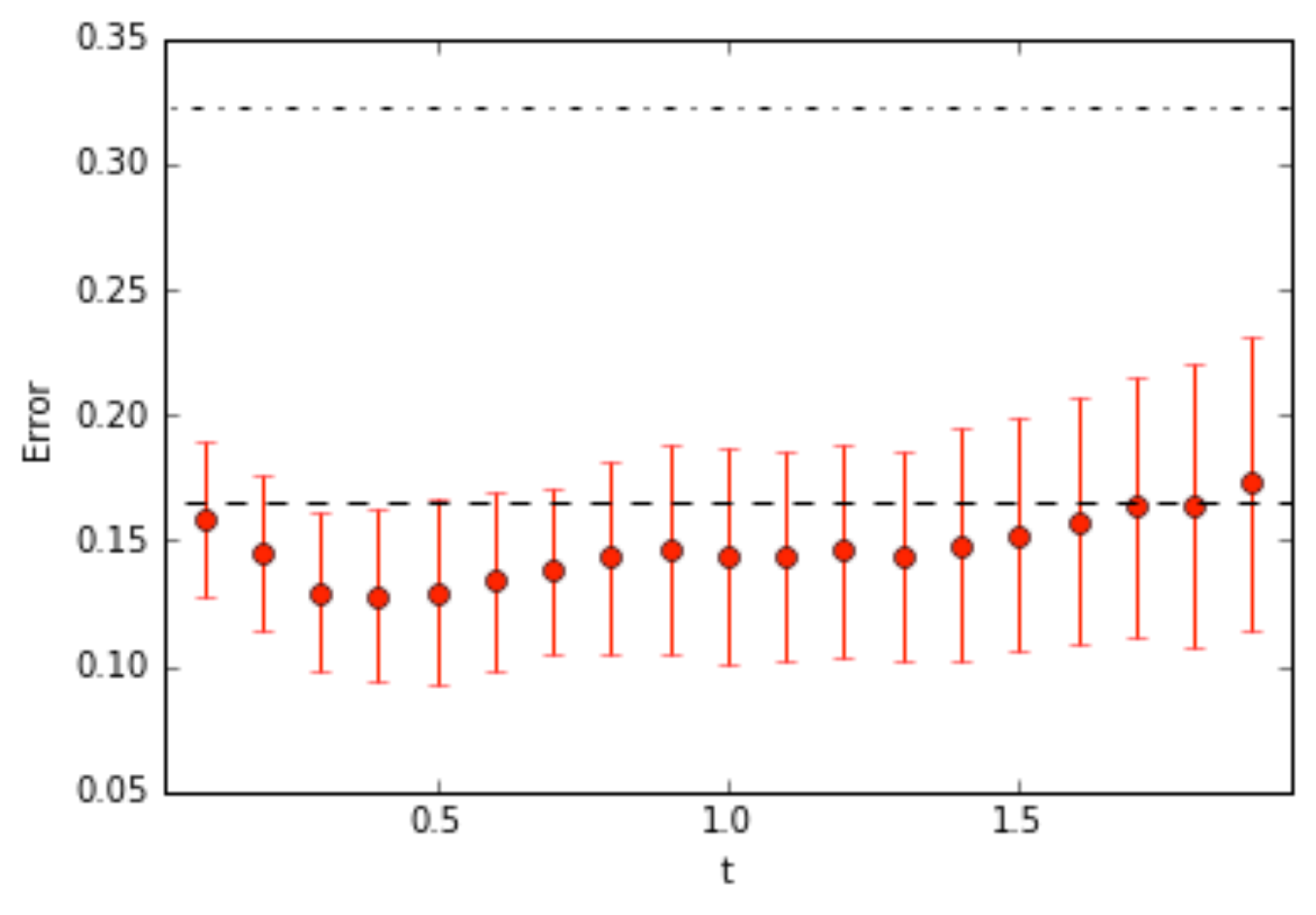}}
\subfigure[]{\label{fig:Bands}
\includegraphics[width=0.484\columnwidth]{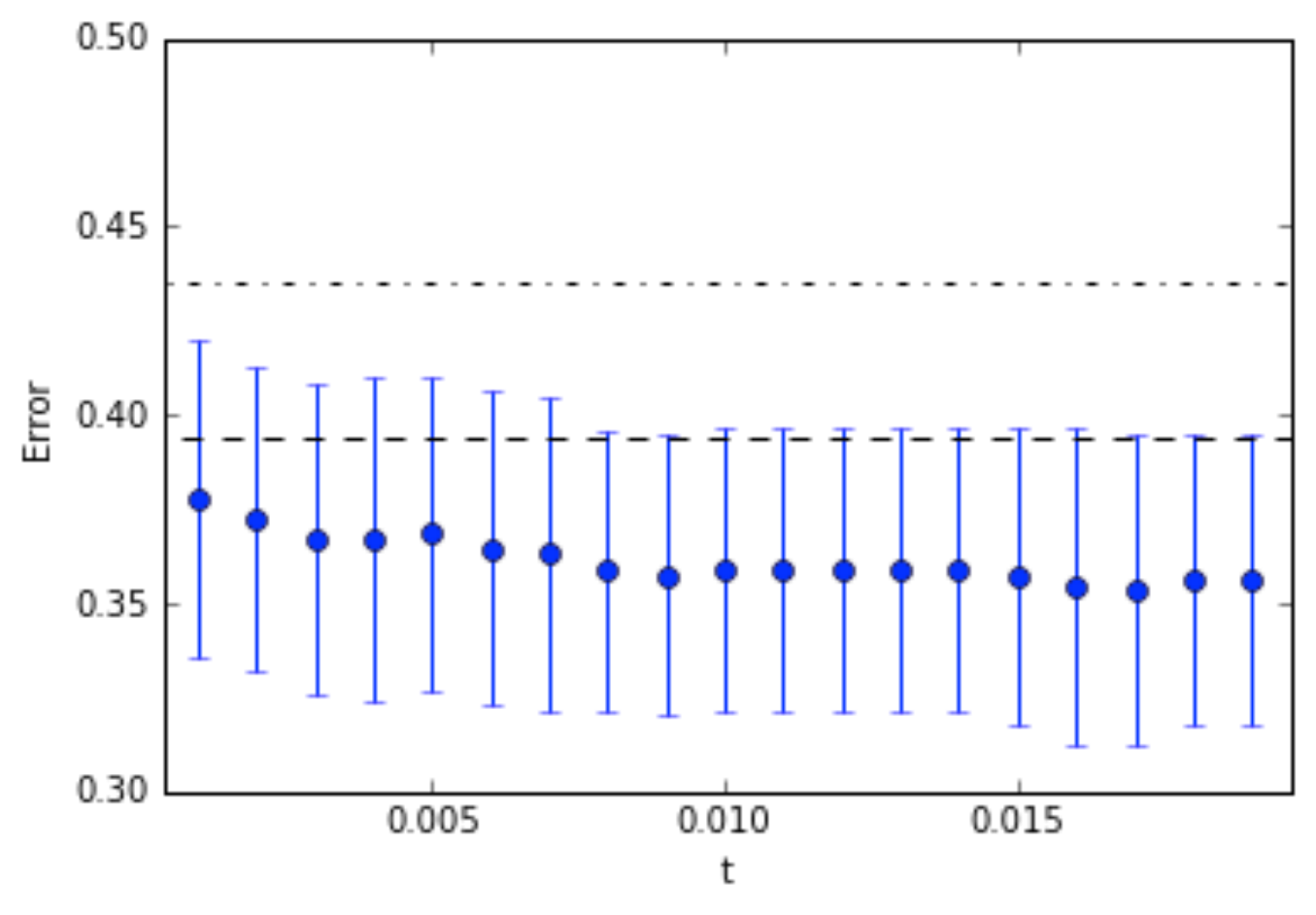}
}
\caption{Comparison between NMC and QNMC performance in terms of the classification error for the datasets (a) \emph{Ionosphere} and (b) \emph{Bands}. In both cases, the simple dashed line represents the QNMC classification error without rescaling, the dashed line with points represents the NMC classification error (which does not depend on the rescaling parameter), points with related error bars (red in (a) and blue in (b)) represent the QNMC classification error for increasing values of the parameter $t$. In (a) $t\in [0.1,1.9]$ and it increases with step $10^{-1}$. In (b), $t\in [0.001,0.019]$ and it increases with step $10^{-3}$.}
\label{fig:compare}
\end{figure}

For any value of the parameter $t$ it can be proved \cite{sergioli2017} that, while the NMC is invariant under rescaling, for the QNMC this invariance does no longer hold.   
Interestingly enough, it is possible to evaluate this interesting property of the QNMC as an advantage for the classification process. In other words, by a suitable choise of the rescaling factor is possible, in principle, to get a decreasing of the classification error. At this purpose, we have studied the variation of the QNMC performance (in particular of the classification error) in terms of the \emph{free} parameter $t$ and in Fig. 1 the results for the datasets \emph{Ionosphere} and \emph{Bands} are shown. In the figure, each point represents the mean value (with related standard deviation represented by the vertical bar) over ten runs of the experiments. We can observe that, for the considered datasets, the QNMC performance for the most of $t$ values is better than the NMC, but for some particular value of $t$ the error gets a further significant reduction (with respect the unrescaled case).

 Let us note that the range of the rescaling parameter $t$ for which the QNMC performance improves, is generally not unique and depends on the dataset. For instance, in Fig. \ref{fig:compare}, we observe that the classification error provided by the QNMC decreases for $t$ ranging from $0.1$ to $1.9$, in the \emph{Ionosphere} case, and from $0.001$ to $0.019$ in the \emph{Bands} case. Then,  we do not generally get an improvement in the classification process for any $t$ ranges. On the contrary, there exist some intervals of the parameter $t$ where the QNMC classification performance is worse than the case without rescaling. Then, each dataset has specific and unique characteristics (in accord to the No Free Lunch Theorem) and the incidence of the non invariance under rescaling in the decresing of the error is, in general,  to determinate by empirical evidences. 
\section{Conclusions and future work}
\label{sec:concluding}
In this work a quantum counterpart of the well known Nearest Mean Classifier has been proposed. We have introduced a quantum minimum distance classifier, called Quantum Nearest Mean Classifier, obtained by defining a suitable encoding of real patterns, \emph{i.e.} \emph{density patterns}, and by recovering the trace distance between density operators. 

We propose a new encoding of real pattern into quantum object that was suggested by recent debates on quantum machine learning according to which, in order to avoid a loss of information caused by encoding a real vector into a quantum state,  we need to normalize the real vector mantaining some information about its norm. Secondly, we defined the \emph{quantum centroid}, \emph{i.e.} the pattern chosen as the prototype of each class, which is not invariant under uniform rescaling of the original dataset (unlike the NMC) and seems to exhibit a kind of  sensitivity to the data dispersion. 

The experiments are organized as follows: both classifiers have been compared in terms of significant statistical indeces. In particular, we considered fourteen different datasets having different nature (real-world and artificial). Further, the no-invariance under rescaling of the QNMC suggested to study the variation of the classification error in terms of a free parameter $t$, whose variation produces a modification of the data dispersion and, consequently, of the classifier performance. In particular we have shown as, in the most of the cases, the QNMC exhibits a significant decreasing of the classification error (and of the other statistical parameters) with respect of the NMC and, for some case, the non invariance under rescaling can provide a significant positive incidence in the classification process.

Let us remark that, even if there is not an absolute superiority of QNMC with respect to the NMC, the method we introduced allows to get some relevant improvements of the classification when we have an \emph{a priori} knowledge about the distribution of the dataset we have to deal with.

In light of such considerations, further developments of the present work will be focused on: \emph{i)} finding out the encoding (from real vectors to density operators) that guarantees the \emph{optimal} improvement (at least for a finite class of datasets) in terms of the classification process accuracy; \emph{ii)} obtain a general method to find the suitable rescaling parameter range to apply to a given dataset in order to get a further improvement of the accuracy; \emph{iii)} understanding for which kind of distribution the QNMC performs better than the NMC. At this purpose, it will be useful to compare the \emph{optimal} QNMC also with other standard classical classifiers.

\begin{acknowledgements}
This work is supported by the Sardinia Region Project ``Time-logical evolution of correlated microscopic system'', LR 7/8/2007 (2015). RAS CRP-55. GMB acknowledges support from CONICET and UNLP (Argentina). 
\end{acknowledgements}

\end{document}